# The design of quaternary eutectic solder by machine learning

Zhenhua Guo[1], Xintong Ren[2], Jiahua Jiang[3], Haoyang Liu[1], Yongjun Huo[1], Xiuchen Zhao[1], K. N. Tu[4], and Yingxia Liu[1*]

1: School of Materials Science and Engineering, Beijing Institute of Technology, Beijing, China

2: Expedia Group, Inc. Seattle, WA, USA

3: School of Information Science and Technology, ShanghaiTech University, Shanghai, China

4: Dept. of Materials Science and Engineering, City University of Hong Kong, Hong Kong

*: yingxia.liu@bit.edu.cn

Abstract：

In mobile microelectronic technology, due to the vertical stacking in advanced packaging, there is a need to develop solder joints with hierarchical melting points, especially below 100 °C.  Usually, we select solders from eutectic binary alloys, however, they are limited in the low temperature range.  While the variety of ternary, quaternary, and high entropy eutectic alloys is sufficient, the selection would be difficult. Therefore, we apply machine learning (ML) algorithms to develop the next generation solders. In this paper, we obtain a Sn-Bi-In-Pb quaternary near eutectic alloy composition from ML model. The eutectic points and the alloy compositions were evaluated and continuously improved by experimental input.  A quaternary eutectic alloy (QEA) with high plasticity and low melting point is obtained, and the actual composition is near the result given by ML. We conclude that the application of ML in solder design has shown the potential to overcome the challenge in searching for the next generation eutectic solders, which will have a broad impact on the industry.

I. Introduction

Due to the COVID-19 virus pandemic, the trend of distance teaching, distance medicine, home office, and on-line meeting has increased greatly the need of advanced consumer electronic products, demanding more function, smaller form factor, larger memory, faster and greater rate of data transmission, cheaper cost, and superb reliability. To sustain the trend, advanced packaging technologies with high complexity and integration, such as vertically stacked packaging-on-packaging (POP) and three-dimensional integrated circuits (3D IC) have already shown their impact to our society [1-3]. In these new packaging structures, multiple reflows of solder joints will be required to achieve the vertical stacking. To avoid the remelting of the solders, there is a need for solders with hierarchical melting points. In addition, as a variety of devices are developing, there is a need for a variety of different melting point solders in corresponding packaging strategies. For example, for bio-medical devices and neuromorphic devices, we need to develop low melting point interconnect materials, since these devices can only sustain a low temperature during manufacturing processes. So far, we haven't known any appropriate solders for these applications.

At present, we have the high temperature solder, Sn70Au alloy, with a melting point of 280 °C and a reflow temperature of about 300 °C. We also have the medium temperature solder SAC 305 alloy with a melting point of 217 °C and a reflow temperature of about 250 °C. For the low temperature solder, we have SnBi eutectic solder with a melting point of 139 °C and a reflow temperature of about 150 °C [4-7]. However, the brittleness of SnBi eutectic solder limits its application in most mobile electronic products. Therefore, it is crucial to develop a solder with a low melting point and appropriate mechanical properties.

The major reason that we need eutectic alloys for solder is to have a single melting point. Thus, thousands of solder joints can melt and solidify at the same time. Therefore, our goal is to search for a low melting point quaternary, or even quintenary eutectic solder. Because the pool of candidates is too big and it's impossible to do experiments of each composition one by one. This is the reason we use machine learning (ML) to solve the problem.

Machine learning can build models and find correlations to predict material properties, including composition, microstructure, physical and chemical properties. Recently, several applications have been successfully applied ML to find new HEA. Huang et al. developed three ML algorithms to predict HEA phase selection and found that atomic size difference and valence electron concentration were two key parameters [8]. Wen et al. designed HEA with ultra-high hardness above 850 HV based on ML prediction [9]. Wu et al. successfully applied ML in designing the quintenary Al-Co-Cr-Fe-Ni high entropy eutectic [10].

In this study, ML was applied to quantitatively determine the composition of QEA. With experimental data, we report a QEA with excellent mechanical properties. The method of finding this QEA can have a wide range of applications.

II. Experimental

2.1 Database and ML model

To build up the database, we introduce the concept of eutectic degree, which means the deviation to the eutectic composition. The smaller the deviation, the higher the eutectic degree. Eutectic degree can be evaluated by primary phase mole fraction and melting point range. The primary phase molar fraction is the percentage of the solid phase solidified before the eutectic reaction. A small primary phase molar fraction means a high eutectic degree. The melting point range is the temperature interval between the beginning and the end of eutectic reaction. We consider the alloy to be eutectic when the melting point range is less than 3 °C.

Primary phase molar fraction and melting range can be obtained from phase diagram. Although the phase diagrams of multi-component alloys with four or more elements are too complicated to be obtained, we can have a "pseudo phase diagram" by the phase diagram calculation (CALPHAD) software JMAT. To apply ML in searching for the proper solder alloy, a database for the Sn-Bi-In-Pb alloy system was established by the CALPHAD calculation of JMAT. The reason that we choose Sn-Bi-In-Pb system is because these four elements are the most common low melting point elements. To ensure the accuracy of ML, a data base containing at least several hundred data sets is needed. In this study, 300 data sets were obtained using JMAT by randomly changing the content of each element in the Sn-Bi-In-Pb system. Each data set contains the elemental composition, the primary phase mole fraction, and the melting point range. The data in the database were split into 70% as training, 15% as testing and 15% as validation. Then the ML models used for predicting QEA with near-eutectic composition were trained. The learning model was constructed based on convolutional neural network (CNN). The architecture of the entire CNN model includes 1D convolution layers to extract features, dropout layers to avoid overfitting of the data, flatten layer and standard fully connected layers. Fig. 1 shows the pictorial elucidation of the CNN model. The input is the percentage of elements for each combination. The output is the melting range and melting temperature.

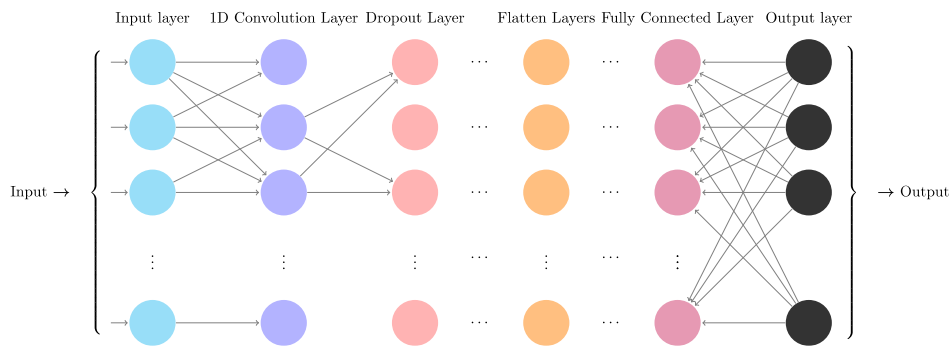

Fig. 1. The pictorial elucidation of the CNN model

To measure the accuracy of the CNN model, we employ the mean absolute error as the loss function. To obtain more accurate combination of the elements, for instance, the optimum combination is not included in the database, we perform a greedy search process for prediction. The specific steps of the greedy search process are provided as follows:

1. use the trained model to predict the $8^4$ points that are uniformly distributed in the $[0, 1]^4$ hypercube;
2. choose 10 points corresponding to the 10 lowest melting point to narrow the search area where the optimum combination is located;
3. randomly 100 points around each one of these 10 points selected in step 2;
4. choose the point that has the smallest melting point among the points sampled in step 3.

Furthermore, to ensure the accuracy of the composition, the result computed by the CNN model was verified and improved by the experimental data.

## 2.2 Property test

In order to verify the results of ML model, we did experiments. We first prepared the samples according to the composition given by ML results. SnBiInPb solder was prepared using high purity (>99.9%) Sn, Bi, In and Pb. The ingots melted completely at around 300 ºC in a vacuum induction furnace under argon atmosphere. After melting, the samples were cooled down to room temperature in air. Pieces about 5-10 mg were cut from the bulk solder alloy. These pieces were analyzed by differential scanning calorimetry (DSC). If the result given by ML is not a perfect eutectic, we will adjust the composition until finding a perfect eutectic with the DSC endothermic and exothermic peaks both having only one peak, which we define it as QEA.

The solder alloy bulk was cut into tiny pieces. These pieces were placed on a hotplate and reflowed on Cu under flux at 100 ºC for different length of time. After soldering, the samples were removed from the hotplate and cooled to room temperature. The samples were mounted using epoxy resin and then cross-sectioned and polished. Besides, original samples without reflow process were made as comparison. The cross-sectional interfaces of the polished samples were observed using scanning electron microscope (SEM). The elemental composition of solder and intermetallic compounds (IMC) were analyzed using energy dispersive X-ray spectroscope (EDX).

After we obtained the QEA from both ML output result and the verifying experiment, we carried out shear test and tensile test to investigate the mechanical properties of the alloy. For the tensile test, QEA solder bulk was processed into tensile samples with length of 10 mm and thickness of 1.5 mm. The tensile samples were tested by Instron 5966 tensile test machine with an extension rate of $1\times10^{-3}$ s$^{-1}$. For the shear test, $4.5 \pm 0.5$ mg of diced QEA solder, Sn58Bi and pure Sn pieces were reflowed on 1 mm diameter circular Cu substrate. Subsequently the shear tests were performed using PTR-1100 shear test machine at room temperature with an elongation rate of 0.05 mm/s.

III. Results

3.1 Melting point measurements

Based on the ML model, a near eutectic composition of sample 1 (Sn-35Bi-47.5In-

4Pb) was obtained. Three samples (sample 2, 3, 4) whose composition are Sn-30Bi-52.5In-4Pb, Sn-40Bi-42.5In-4Pb and Sn-37.5Bi-45In-4Pb, respectively, were made to search for a perfect eutectic composition. Fig. 2 (a) shows the strategy in composition modification. On the basis of sample 1, the content of Bi decreased by 5% and the content of In increased by 5%, we obtained sample 2. Then the content of In decreased by 5% and the content of Bi increased by 5%, we obtained sample 3.

To confirm the results given by ML, we did the DSC tests for the samples, as shown in Fig. 2(b) to (e). Fig. 2(b) shows that the ML sample has only one endothermic peak, and two exothermic peaks in the DSC curve, indicating that the alloy is almost eutectic, but there is phase transformation during the solidification. In Fig. 2(c), the endothermic peak of sample 2 is irregular, and there are more exothermic peaks than sample 1, showing that the composition adjustment in this direction does not meet our requirements. In Fig. 2(d), the endothermic peak of sample 3 is multiple, but the exothermic peak becomes single, just contrary to sample 1. Therefore, we believe the eutectic composition should be between the composition of sample 1 and sample 3. As a result, we obtained sample 4. As shown in Fig. 2(e), sample 4 has only one peak for both exothermic and endothermic process, so it can be considered as an QEA alloy. The melting point of this alloy is 60.81 ºC.

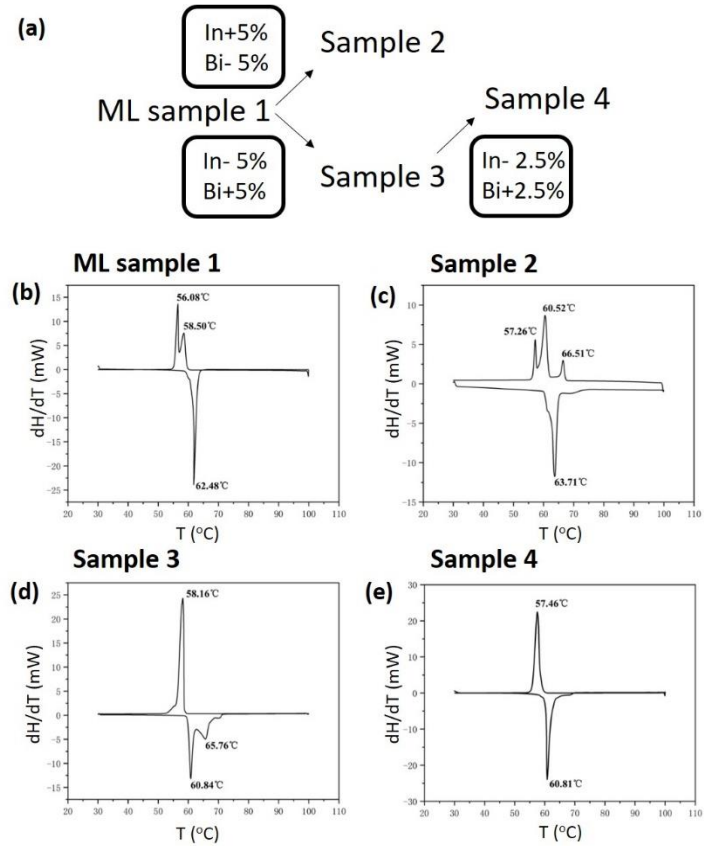

Fig. 2. (a) Illustration on the composition modification. The DSC curve of (b) sample 1, (c) sample 2, (d) sample 3, and (e) sample 4.

3.2 Microstructure characterization

Sample 2 is far from the desired eutectic composition, so we didn't do further investigation on that sample. Fig. 3 shows the SEM pictures of original sample 1, 3 and 4, which didn't go through reflow. The lamellar eutectic structure can be observed. There are mainly two phases in the three alloys, and we call them light phase and dark phase. Distinctively, three phases, including a darker phase can be observed (circled in red) in Fig. 3(d) of sample 3(Sn-40Bi-42.5In-4Pb).

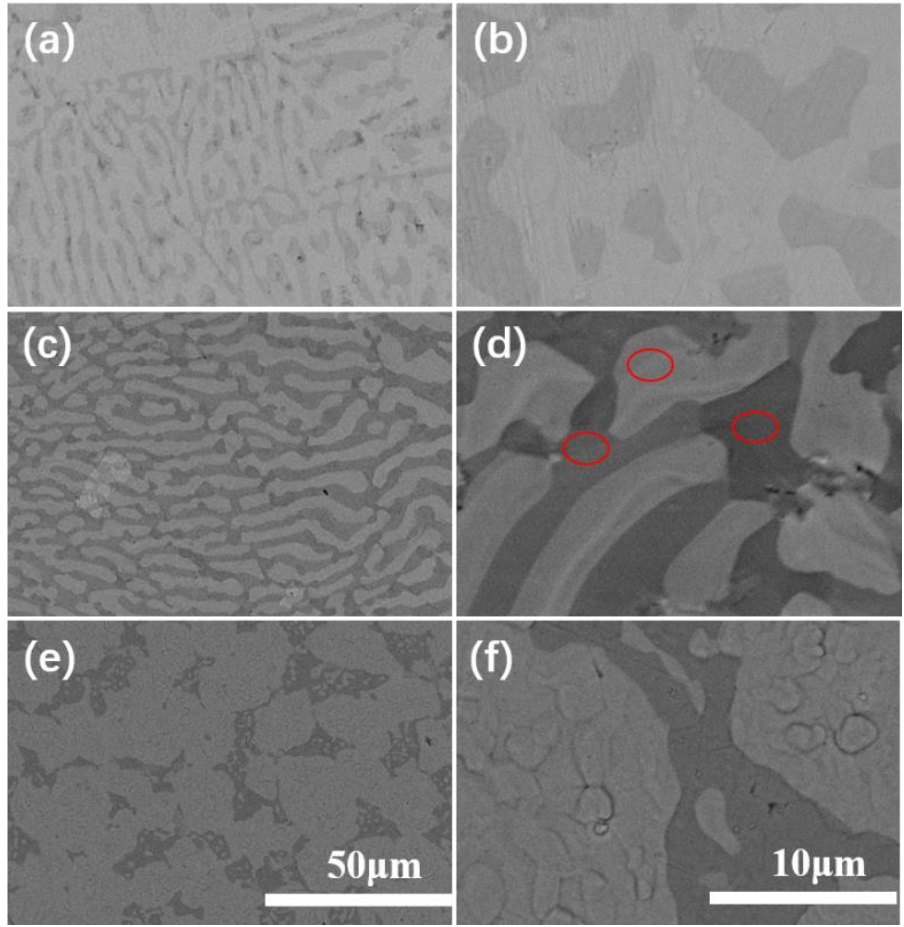

Fig. 3. The original sample microstructure of (a), (b) sample 1 (Sn-35Bi-47.5In-4Pb), (c), (d) sample 3 (Sn-40Bi-42.5In-4Pb), and (e), (f) sample 4 (Sn-37.5Bi-45In-4Pb).

Fig. 4 shows the SEM pictures of sample 1, 3 and 4 reflowed at 100 ºC on Cu substrates for 30 min. The wetting angles measured from the SEM micrograph for sample 1, 3 and 4 are 26.5°, 28.4°, and 28.0°, respectively. Similar to the original sample, there are mainly two phases in these alloys. But a darker phase can be found (circled in red) in Fig. 4(b) of sample 1(Sn-35Bi-47.5In-4Pb).

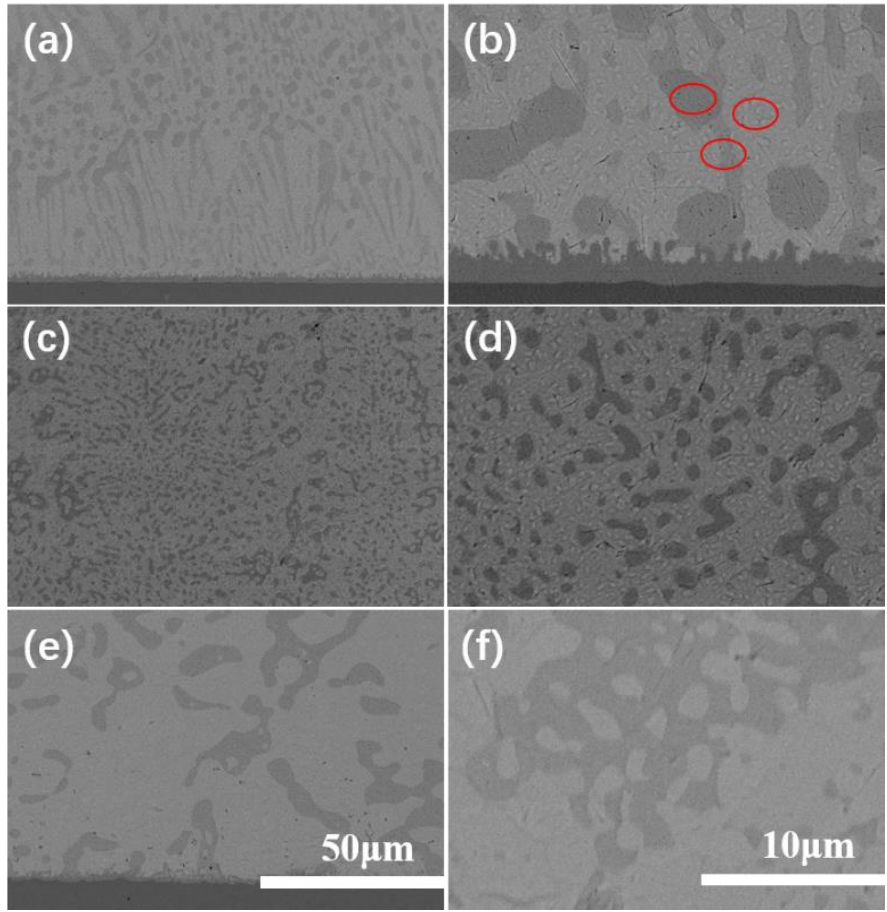

Fig. 4. The reflowed sample microstructure of (a), (b) sample 1 (Sn-35Bi-47.5In-4Pb), (c), (d) sample 3 (Sn-40Bi-42.5In-4Pb), and (e), (f) sample 4 (Sn-37.5Bi-45In-4Pb).

Fig. 5 shows the EDX mapping result of the eutectic solder (sample 4) after reflow. There are mainly two phases, Bi-rich phase, and Sn-rich phase. In and Pb atoms are evenly distributed in the sample. Table 1 shows the EDX results of each phase in the samples. We also did XRD to further confirm the EDX results, as shown in Fig. 6. The XRD result is complicated to analyze, and there are some weak peaks can't be indexed, which we tend to believe they are from the Sn-rich phase. We will discuss this point in Section 4.2. According to the EDX and XRD results, $BiIn_2$ and $Bi_3In_5$ are the main components in Bi-rich phase, with some Sn and Pb atoms substituting In and Bi atoms.

The original sample 1 has the BiIn$_2$ phase, because the content of In is the largest in sample 1, and that phase changes to two phases of BiIn$_2$ and Bi$_3$In$_5$ after reflow. The original sample 3 has the Bi$_3$In$_5$ phase, and two Sn-rich phases before reflow. Sample 4 seems to have stable Bi$_3$In$_5$ and Sn-rich phases, which don't go through phase transformation during reflow.

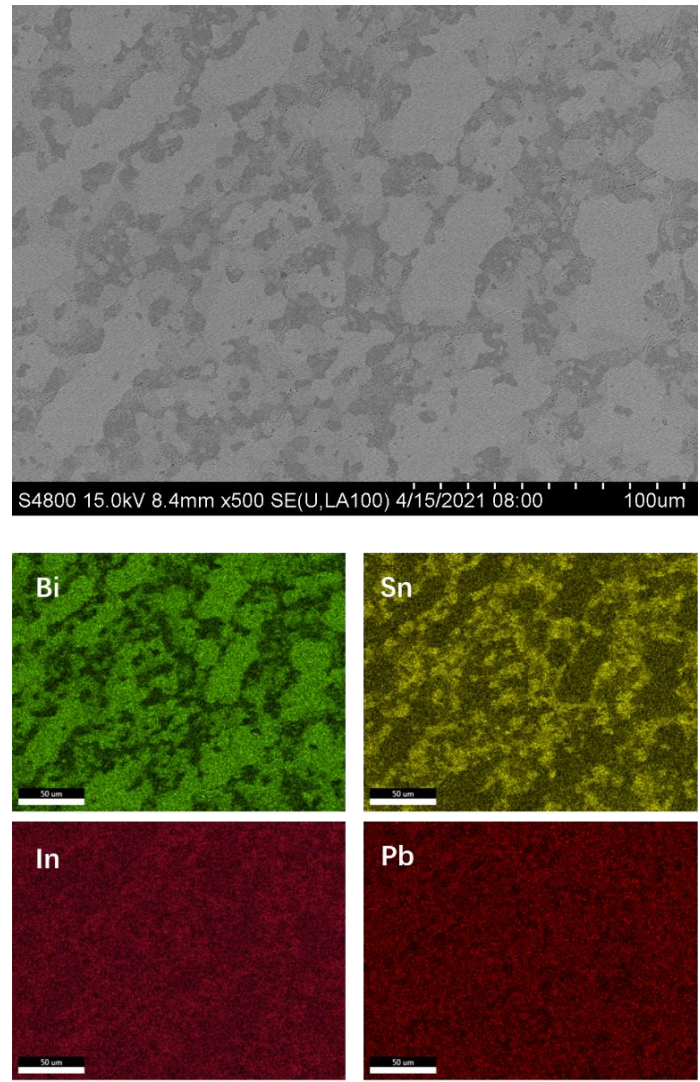

Fig. 5 EDX result of the eutectic solder sample 4 after reflow.

Tab. 1. The EDX results of sample 1, 3, and 4 before and after reflow.

| | BiIn phase (at. %) | | SnIn phase (at. %) | |
|---|---|---|---|---|
| Original-1 | Sn: 23.2  Bi: 13.7  In: 58.9  Pb: 4.2 | | Sn: 59.5  Bi: 3.3  In: 33.5  Pb: 3.7 | |
| Reflow-1 | Sn: 20.7  Bi: 14.0  In: 60.6  Pb: 4.7 | Sn: 9.6  Bi: 16.7  In: 68.3  Pb: 5.4 | Sn: 59.7  Bi: 4.6  In: 33.4  Pb: 2.3 | |
| Original-3 | Sn: 17.6  Bi: 34.3  In: 47.7  Pb: 0.4 | | Sn: 49.3  Bi: 7.3  In: 41.7  Pb: 1.7 | Sn: 25.4  Bi: 6.3  In: 63.6  Pb: 4.7 |
| Reflow-3 | Sn: 17.8  Bi: 34.4  In: 47.4  Pb: 0.4 | | Sn: 72.6  Bi: 8.7  In: 17.3  Pb: 1.4 | |
| Original-4 | Sn: 17.9  Bi: 30.8  In: 47.0  Pb: 4.3 | | Sn: 60.1  Bi: 2.9  In: 35.0  Pb: 2.0 | |
| Reflow-4 | Sn: 15.9  Bi: 35.3  In: 48.4  Pb: 0.4 | | Sn: 54.7  Bi: 5.3  In: 37.7  Pb: 2.3 | |

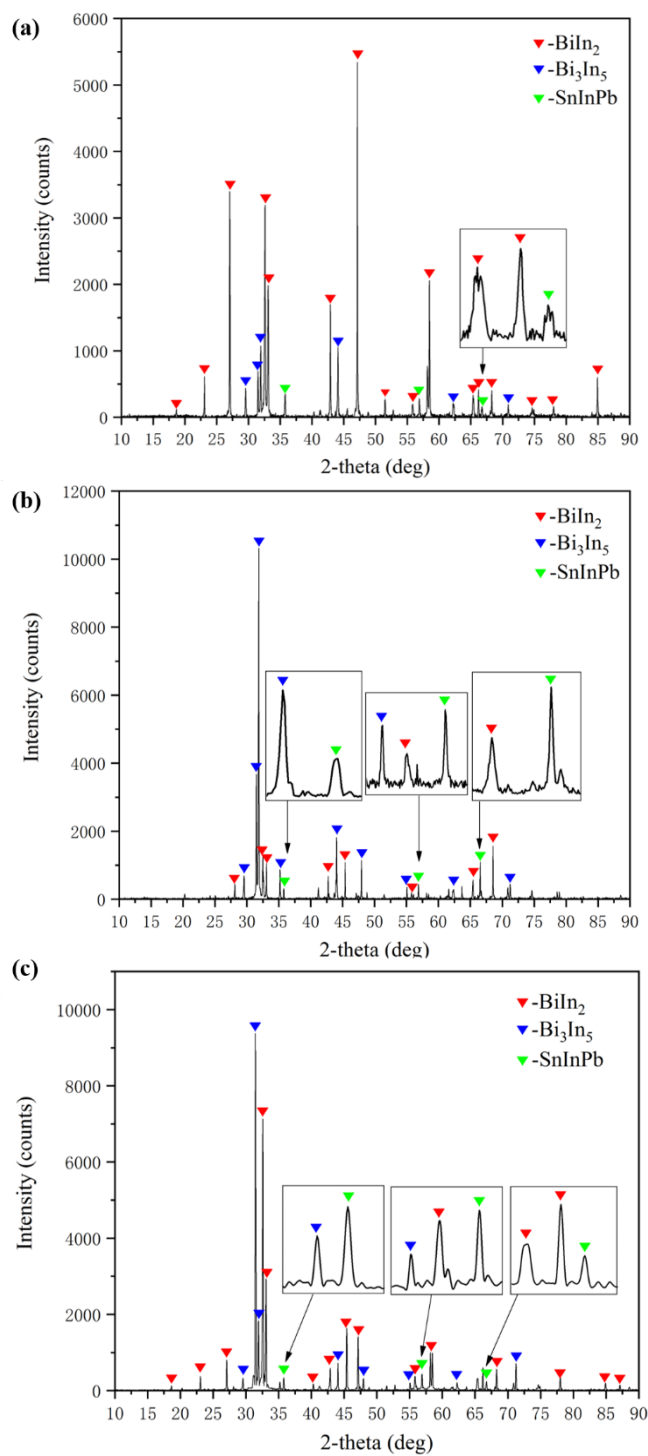

Fig. 6. The XRD spectrum of (a) sample 1, (b) sample 3, and (c) sample 4 after reflow.

We did TEM to investigate the Sn-rich phase, as shown in Fig.7. Fig. 7(b) and (c)

show the electron diffraction pattern of sample 4, the corresponding areas are circled in Fig. 7(a). The indexed result indicates that the phase has FCC structure, with the average lattice parameter as a = b = c = 0.478 nm. The TEM-EDX result show that the sample is mainly a mixing structure of Sn and In.

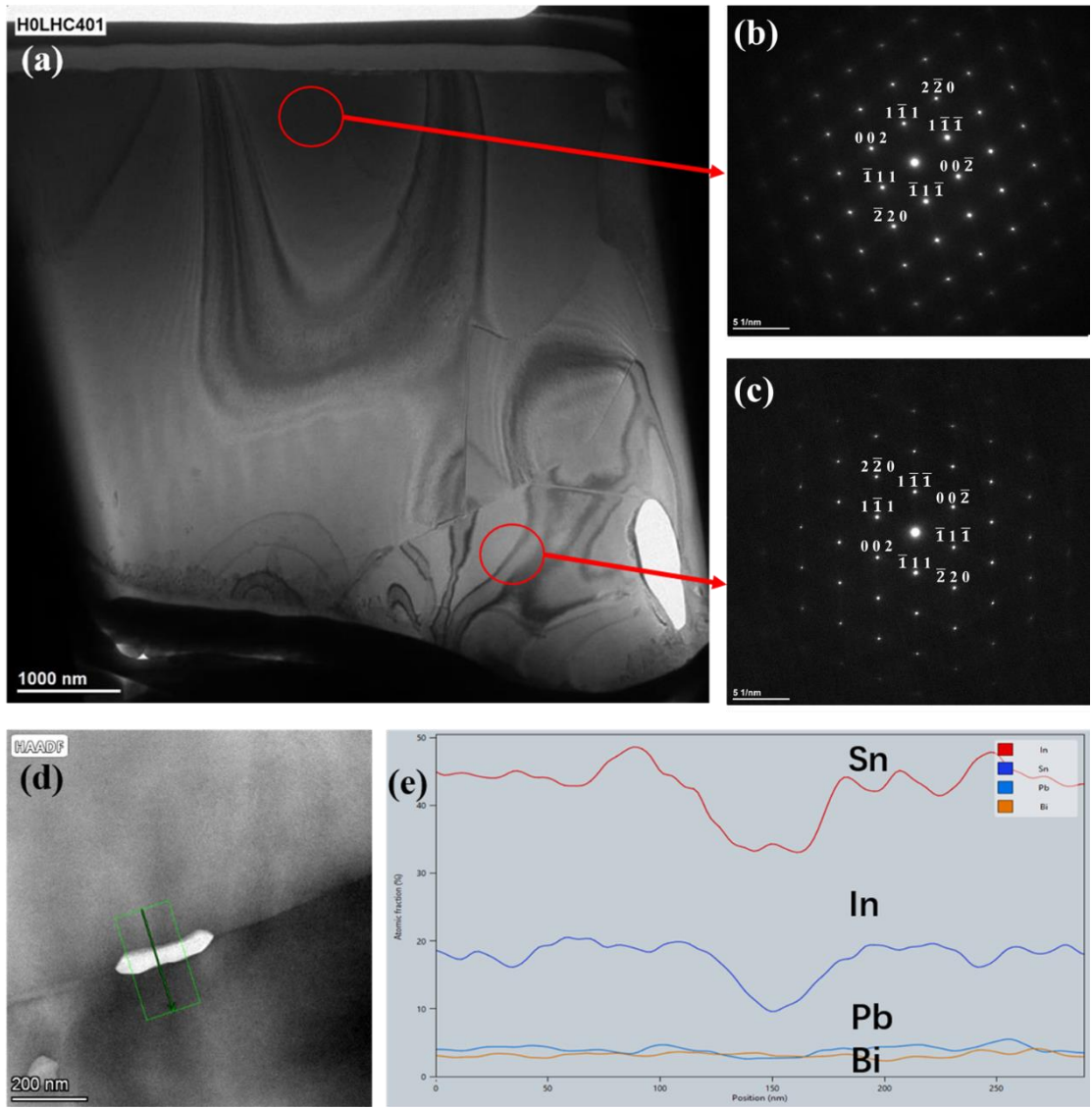

Fig. 7. The TEM diffraction pattern of sample 4 (Sn-37.5Bi-45In-4Pb).

3.3 Mechanical properties

The tensile test results are shown in Fig. 8, and the average elongations of samples

1, 3, and 4 are 33.3 ± 5.2%, 40.1 ± 10.4% and 117 ± 7.5%, respectively, and the average tensile strength is 22.45 ± 2.07 MPa, 28.13 ± 1.08 MPa and 22.55 ± 0.59 MPa, respectively. Sn58Bi and SAC 305 solder tensile stress data were collected from published work for comparison. The tensile strength of Sn58Bi and SAC 305 solder is around 60 MPa and 40 MPa, respectively. The elongations of Sn58Bi and SAC 305 are similar, around 30 to 50% [11-14]. Compared with Sn58Bi and SAC 305 solder, the QEA solder has a lower tensile strength and a notably higher elongation. But the other two quaternary alloys have an elongation similar to Sn58Bi and SAC 305 solder. The samples after tensile fractured are shown in Fig. 9. Different from the other four samples, a tip can be observed for the QEA sample after tensile fracturing.

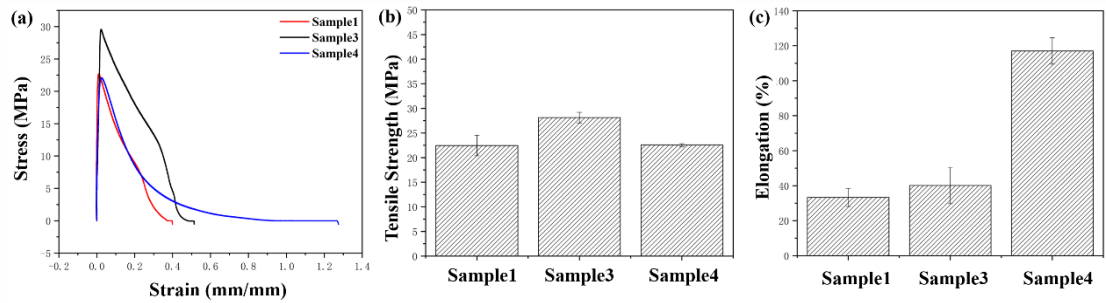

Fig. 8. Tensile test results for sample 1, 3, and 4, (a) the tensile curve, (b) the tensile strength, and (c) the elongation.

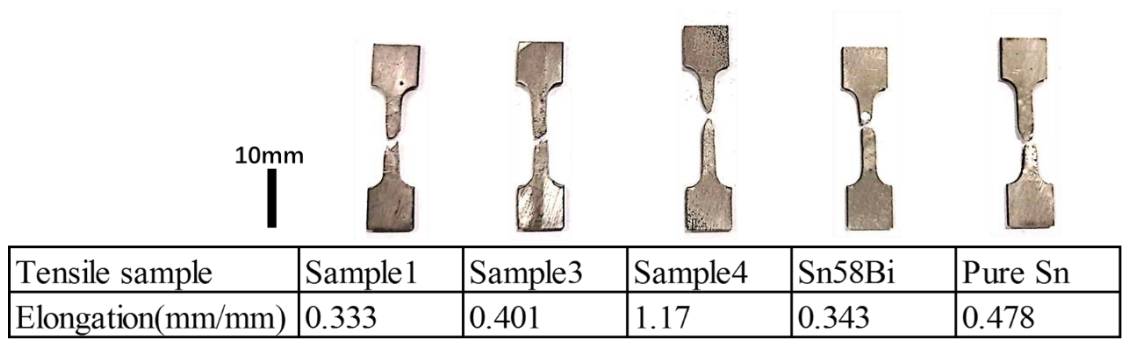

| Tensile sample | Sample1 | Sample3 | Sample4 | Sn58Bi | Pure Sn |
|---|---|---|---|---|---|
| Elongation(mm/mm) | 0.333 | 0.401 | 1.17 | 0.343 | 0.478 |

Fig. 9. The fractured surface of samples after tensile tests.

We also did the shear test. The shear strength of eutectic sample 4, Sn58Bi and pure

Sn solder joints is measured as 26.14 ± 3.71 MPa, 43.65 ± 2.52 MPa and 48.32 ± 1.74 MPa, respectively. The measured shear strength of Sn58Bi and Sn solder joints is in the range of reported data [15-19]. Compared with Sn58Bi and Sn solder joint, the QEA solder has a lower shear strength.

IV. Discussion

4.1 Microstructure and melting point

With the DSC and the microstructure characterization results by SEM and XRD, we have a deeper understanding in the relationship between the microstructure and melting point. There are two endothermic peaks in sample 1 and two exothermic peaks in sample 3, which are due to the formation of a new phase. In sample 1, due to the excess of In atoms, the originally formed phase is $BiIn_2$, and after reflow, it transforms to $BiIn_2$ and $Bi_3In_5$. In sample 3, due to the excess of Bi atoms, the originally formed phase is $Bi_3In_5$. In addition to that phase, there is some high-In Sn-rich and low-In Sn-rich. After reflow, the high-In phase and low-In phase merges to a more stable Sn-rich phase. Compared to sample 1 and sample 3, sample 4 has one endothermic peak and one exothermic peak, with only 3 °C melting range, which can be regard as a eutectic alloy. The alloy has stable $Bi_3In_5$ and Sn-rich phase, both in melting and solidification processes. Thus, in the eutectic structure, the two phases are $Bi_3In_5$ and Sn-rich phase.

The Sn-rich phase has been investigated and is believed to be a high entropy phase.

In the XRD result, in addition to the peaks from $BiIn_2$ and $Bi_3In_5$, we have some peaks that are unable to be indexed. The three main peaks are $2\theta = 35.78°$, $57.14°$, and $66.75°$. According to the TEM result, the phase is mainly a mixture of Sn and In, with FCC lattice structure, and the lattice parameter is around 0.478 nm. The measured atom radius would be 169 pm. Because the EDX result shows that the phase only has Sn, In and Pb atoms, and the atomic radius for Sn, In, and Pb are 140 pm, 162 pm, and 175 pm, respectively, so the measured 169 pm is among them, indicating it's reasonable to believe the phase is a mixture of Sn, In, and Pb atoms. With the measured lattice parameter of 0.478 nm, we calculated the corresponding XRD $2\theta$ values for FCC (111), (200), (220), and (311), which are 32.38°, 37.56°, 54.18°, and 64.54°. Compared with the three unknown XRD peaks, the measured values are around 2-3° higher than the calculated ones, which is reasonable. The absence of (200) peak may be due to the overlap of other peaks. Together with the XRD and TEM results, we believe the eutectic alloy has the SnBiInPb high entropy phase.

One interesting point is that, the undercooling of this eutectic phase in solidification is around 3 °C. Compared with SAC 305, Sn37Pb and Sn58Bi, the reported undercooling for the alloys is about 10 to 20 °C [20-22], and we can conclude the undercooling of the QEA is quite small, implying the nucleation of the two phases in the eutectic alloy is easy. $Bi_3In_5$ is an IMC phase, with a melting point around 88 °C [23]. The high entropy phase of SnBiInPb, may reduce the nucleation barrier, and thus reduce the undercooling. Because of the small undercooling, the eutectic structure is quite different from the traditional eutectic lamellar structure. As shown in Fig. 3 (e) and (f), the eutectic microstructure is "cloud-like", rather than "lamellar-like". This unique eutectic structure may contribute to the high plasticity that we are going to discuss below.

4.2 High plasticity of the QEA

As shown in Fig. 8(a) and (b), the QEA sample (sample 4) exhibits extraordinary plasticity, compared with the other four alloys. According to the tensile test curve, the samples exhibit work softening rather than work hardening as they deform, since the deformation is occurring at high homologous temperatures of SnIn alloy, at which recovery and recrystallization should dominate over hardening.

However, we noted that sample 1 and sample 3 also have a low melting point, around 60 ºC, but their elongation is much lower than QEA. One possible explanation is that, the QEA has a very small undercooling, leading to the "cloud-like" eutectic microstructure, rather than lamellar. For sample 1 and sample 3, because they are not perfect eutectic, and the undercooling is larger than 5 ºC, they have lamellar microstructure. The lamellar microstructure should have less sliding mobility than "cloud-like" microstructure, due to anisotropic. Another possible reason is that the QEA sample has more weight of SnBiInPb high entropy phase, so that it can play a key role in the whole alloy. A detailed research on the high elongation of QEA will be carried out later.

The small undercooling of QEA is meaningful in the design of solders.  This is because a large undercooling means a large nucleation barrier, which is undesirable.

For a small undercooling, the difference is small on which phase nucleates first. Technically, the advantage of small undercooling is that we can have all the solder joints solidified almost at the same time. Another advantage is that the relationship between small undercooling, cloud-like microstructure, and plasticity may help us to improve the ductility of solder alloy in the application of mobile devices.

4.3 ML model in solder design

In this work, our data base is not big enough so far, and we have only 300 data in the set of Sn, Bi, In and Pb. The result given by ML is not perfect, however, it's quite near our goal. By trials with limited times, we can find the perfect eutectic solder alloy. Although this eutectic SnBiInPb solder may be too soft in solder application, nevertheless, it's promising to be applied to the solid-liquid diffusion bonding process [24, 25].

We are now working on a much bigger data base, and the developed ML model can predict the eutectic composition from almost any Sn-based element combinations. We are confident that ML can overcome the challenge of searching for the future eutectic solders, which seems to be impossible by experimental search in the old days.

V. Conclusion

We applied ML to the search for a new QEA. The composition given by the ML model is Sn-35Bi-47.5In-4Pb. Experimental data confirm that this alloy is near eutectic. By modifying the composition to Sn-37.5Bi-45In-4Pb, we obtained a QEA.

This alloy has tensile strength of 22.55 MPa and elongation of 1.17. The high elongation is explained by the "cloud-like" microstructure. This soft and low melting point QEA may be applied to the solid-liquid diffusion bonding in packaging technology. This work has shown the potential to design the future solder alloys by ML.

Acknowledgements

This work was supported by Young Scientists Fund of National Natural Science Foundation of China [grant number 51901022].